\documentclass[12pt]{article}

\usepackage{cite}

\usepackage{graphicx}
\usepackage{times}
\usepackage{amsmath}
\usepackage{siunitx}
\usepackage[usenames,dvipsnames]{xcolor}

\usepackage{lineno}

\newcommand{\cecilia}[1]{{\color{red}#1}}

\newcommand{\maurizio}[1]{{\color{magenta}#1}}

\title{Seismic monitoring using the telecom fiber network}

\author
{Simone Donadello$^{1}$, 
Cecilia Clivati$^{1,*}$,
Aladino Govoni$^2$,
Lucia Margheriti$^2$,\\ Maurizio Vassallo$^3$,
Daniele Brenda$^4$, Marianna Hovsepyan$^4$,\\ 
Elio K. Bertacco$^{1}$,  Roberto Concas$^{1}$, Filippo Levi$^{1}$, Alberto Mura$^{1}$,\\ André Herrero$^5$, Francesco Carpentieri $^4$, Davide Calonico$^{1}$\\ 
\normalsize{$^{1}$Istituto Nazionale di Ricerca Metrologica,  strada delle cacce 91, 10135 Torino, Italy}\\
\normalsize{$^2$Istituto Nazionale di Geofisica e Vulcanologia, Sezione ONT, Roma,  Italy}\\
\normalsize{$^3$Istituto Nazionale di Geofisica e Vulcanologia, Sezione Roma1, L'Aquila,  Italy}\\
\normalsize{$^4$ OPEN FIBER, Via Laurentina 449 - 00142 Roma}
\\
\normalsize{$^5$Istituto Nazionale di Geofisica e Vulcanologia, Sezione Roma1, Roma,  Italy}\\
\normalsize{$^\ast$To whom correspondence should be addressed; E-mail:  c.clivati@inrim.it}
}

\date{}

\begin{document} 


\baselineskip24pt

\maketitle 

\section*{Summary}
\textbf{Laser interferometry enables to remotely measure microscopical length changes of deployed telecommunication cables originating from earthquakes \cite{marra2018}.
Long range and compatibility with  data traffic make it unique to the exploration of remote regions \cite{marra2022}, as well as highly-populated areas where optical networks are pervasive, and its large-scale implementation is attractive for both  Earth scientists \cite{lindsey2017,wang2018,jousset2018,ajo2019,lindsey2019,matsumoto2021, currenti2021,flores2023} and telecom operators. However, validation and modeling of its response and sensitivity are still at an early stage \cite{fichtner2022,noe2023} and suffer from lack of statistically-significant event catalogs and limited availability of co-located seismometers. 
We implemented laser interferometry on a land-based telecommunication cable and analyzed 1.5 years of continuous acquisition, with successful detections of events in a broad range of magnitudes, including very weak ones. By comparing fiber and seismometer recordings we determined relations between a cable’s detection probability and the magnitude and distance of events, and showed that spectral analysis of recorded data allows considerations on the earthquake dynamics. Our results reveal that quantitative analysis is possible for this sensing technique and  support the interpretation of data from the growing amount of interferometric deployments.
We anticipate the high integration and scalability of laser interferometry into existing telecommunication grids to be useful for the daily seismicity monitoring,  in perspective exploitable for civilian protection use. 
}

 Probing length changes of deployed telecommunication fibers as a way to access ground motion has attracted growing interest from Earth scientists \cite{lindsey2017,wang2018,jousset2018,ajo2019,lindsey2019,matsumoto2021, currenti2021,flores2023} and network operators \cite{zhan2021,ip2021,huang2020}, that foresee the possibility to implement distributed, spatial-aliasing-free sensor grids exploiting the infrastructure already in place for global telecommunications in a scalable and sustainable way. While traditional techniques  \cite{zhan2020a} have a short range and are not compatible with standard data transmission, novel approaches based on coherent laser interferometry \cite{marra2018, marra2022} and state-of-polarization sensing \cite{zhan2021} enabled earthquake detection on telecommunication cables laid in the ocean floors \cite{marra2022,zhan2021,mecozzi2021} and in land \cite{bowden2022,bogris2022,noe2023}. 
Laser interferometry measures the variation of the optical path length experienced by a coherent laser radiation as it travels a fiber subjected to strain \cite{marra2018,bowden2022,noe2023,marra2022}, with higher sensitivity and broader linear range compared to polarization-based sensing \cite{mecozzi2021}. Its implementation on land-based fibers allows more insightful interpretation  of the fiber response than with sub-sea cables, thanks to a larger number of traditional sensors in cables surroundings  \cite{fichtner2022,bowden2022,noe2023}. However, laser interferometry on land cables is also attractive per se, for monitoring densely-populated areas: here, the fiber pervasiveness compensates for the poor coverage by traditional seismic networks which privilege sensors installation in quieter regions,  and enables applications such as people flows and vehicle traffic monitoring.

We describe the realization of a seismic observatory based on laser interferometry, operational since two years on a telecommunication link in Italy, and present first results of continuous acquisition over a long period, in which we systematically compared  fiber data to those from nearby stations of the Italian permanent seismic network (IV network) managed by Istituto Nazionale di Geofisica e Vulcanologia (see Methods). We found recurrent behaviour and underlying scaling rules, which enabled us to make quantitative  considerations on the cable  response to different classes of events and  application of laser interferometry both as a research and a survey tool in different contexts. 

\begin{figure}
    \centering
    \includegraphics[width=0.9\textwidth]{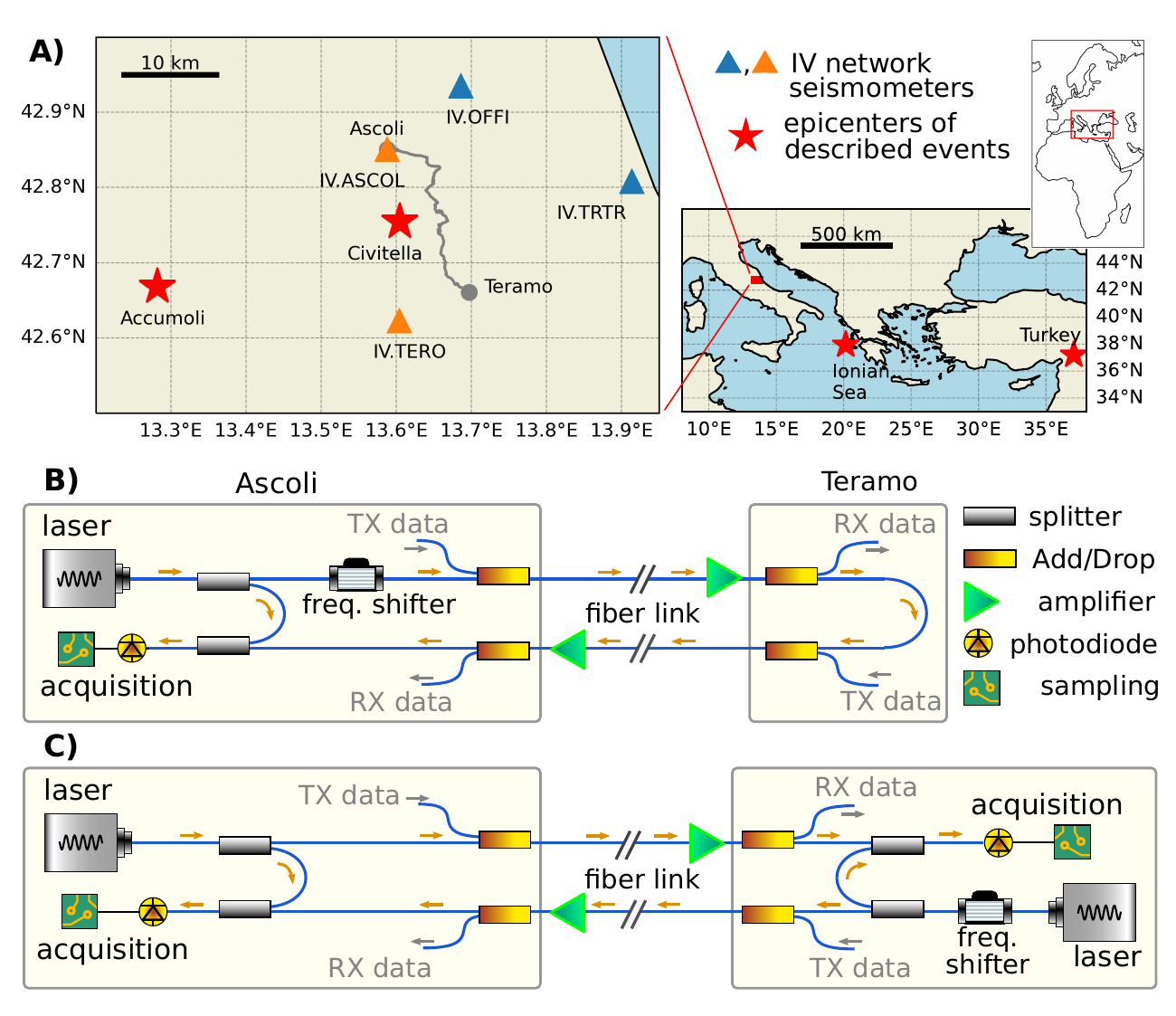}    
    \caption{\textbf{experimental layout.} A) Map \cite{cartopy} of the deployed observatory and its geographical location. Triangles indicate the location of nearby seismometric stations, the closest of which, IV.ASCOL and IV.TERO (orange), were used for comparison throughout the analysis. IV.ASCOL was installed at the fiber end in Ascoli, IV.TERO is about \SI{9}{km} west to the fiber. Stars indicate epicenters of events shown in this study.   B) Sketch of the optical scheme. In Ascoli, the laser radiation is split into two parts: one serves as a reference beam, the other passes through a frequency shifter, is spectrally-combined with telecom data (TX/RX) using optical Add/Drop multiplexers, amplified and launched towards the remote end in Teramo. Here, it is reflected back along an adjacent fiber and, once back in Ascoli,  is recombined to the reference beam to extract an interference signal. C) Modified optical scheme, in which two separate  lasers, hosted at opposite cable ends, are launched in opposite directions and interfered with the local laser. Information about the cable deformations is retrieved by combining synchronous recordings gathered at the two cable ends. The frequency shifter is here  used to add a  periodical pattern to the carrier frequency for synchronizing remote acquisitions.  
    \label{fig:setup}}
\end{figure}

The fiber is  \SI{36}{km} long and is part of a regional ring architecture at \SI{100}{Gb/s} with quadrature phase-shift keying (QPSK) modulation. It is mostly hosted inside a road infrastructure, except for  $\sim$\SI{100}{m} in aerial cable, crosses few bridges and terminates in medium-size towns, thus suffering from diverse and time-varying noise processes. Its average azimuthal direction is \ang{155} (see Extended Data Fig. 1).
Figure \ref{fig:setup}A shows the cable layout and the location of nearest sensors of the IV network. Specifically, IV.ASCOL was installed at one cable end for this experiment and was used, together with IV.TERO, for comparison  throughout the analysis (see Methods). 
The sensing light signal is produced by a compact narrow-linewidth laser, wavelength-multiplexed to the internet traffic and launched into the network. We measure the interference between the signal travelling the round-trip and a local reference beam (Fig. \ref{fig:setup}B) or between two remote laser beams launched in opposite direction  (Fig. \ref{fig:setup}C), in a conceptually similar setup modified in technical details (see Methods) \cite{donadello2023}.

\section*{Results}
\begin{figure}
    \centering
    \includegraphics[width=0.48\textwidth]{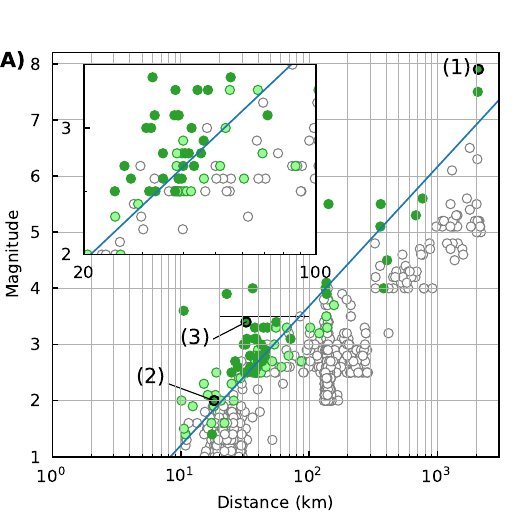}
    \includegraphics[width=0.48\textwidth]{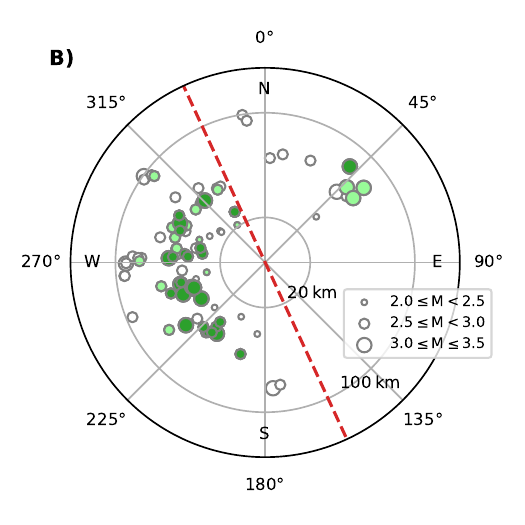}
    \caption{\textbf{Sensitivity of the fiber as an earthquake detector.} A) Map of events considered in the 1.5-year-long catalogue as a function of their magnitude and distance to the fiber. Color indicates if the event was detected (green), unsure (light-green) or non-detected (empty) by the fiber. The line shows the estimated sensitivity threshold for our sensor. Labels (1), (2) and (3) indicate specific events which are described in this study. The inset zooms on events in a range of \SI{20}{km} to \SI{100}{km} and magnitude 2 to 3.5. B) Same events as the inset, mapped as a function of their angular distribution and distance to the fiber. Circles dimensions indicate magnitude class; the dashed line indicates the average direction of the cable.  }
    \label{fig:DM}
\end{figure}

On a standard seismometer, the probability  $P$ to detect an event depends on its magnitude $M$ and hypocenter distance to the fiber $d$. Assuming $P$ is related to the  recorded peak value of a physical quantity (e.g. ground acceleration or velocity), the following relation holds \cite{sabetta1987}:
\begin{equation}
\label{eq:detectionprobability}
P = A_2 \frac{10^{\ A_1 M}}{d}    
\end{equation}
where $A_1$ depends on the type of physical measure and $A_2$ is a generic scaling coefficient. 
For interferometry-based earthquake detection, no scaling law was reported so far. 
We performed this analysis, generating an earthquake catalog that included over 600 events with magnitude between 1.5 and 8 and epicentre distances up to  \SI{2000}{km},  spanning a time frame of 1.5 years. We included all earthquakes located by INGV monitoring system in the considered period and looked for signature of the events on fiber recordings (see Methods). The distribution of 
 detected, unsure and non-detected events as a function of their magnitude and distance to the closest point on the fiber (Fig. \ref{fig:DM}) well agrees with Eq. \ref{eq:detectionprobability}.

The sensitivity threshold line (blue curve in Fig. \ref{fig:DM}) is found iteratively  by  minimizing the cumulative number of detected events below the line and non-detected events above the line and corresponds to $A_1 = 0.403 \pm 0.001$ and $A_2=(3280 \pm 10)$\SI{}{km} (see Methods). Although the fiber suffers
from a higher level of integrated noise as compared to a point-like sensor, $A_1$ is of the same order of magnitude as for IV.ASCOL (see Extended Data Fig. 2) and comparable to the values published in the literature for scaling peak ground accelerations or velocities.

With the better statistics enabled by ongoing data collection,  this estimation could be refined accounting for effects such as the cable length or the coupling between $M$ and $d$, among others.

The fiber shows the capability to detect events with local magnitude $M_l$ as small as 1.6 in the local area and a good detection predictability on events with $M_l>2.5$, in spite of its unfavorable environment. For instance, the background noise recorded by the fiber changes by  \SI{10}{dB} between day and night and shows signature of randomly distributed, non-stationary events which mostly affect detection capability for weakest earthquakes on daily hours. In addition, the two  aerial segments of the cable introduce noise in the spectral region around \SI{1}{Hz}, which is relevant to seismic detection (see Extended Data Fig. 3 and 4). 
While most of these impairments cannot be avoided on telecom-integrated sensing systems, we expect their impact to be less detrimental in  structured architectures based on separated cables    simultaneously probed by the same laser interrogator.
 
We also analysed the detection probability of average-size earthquakes on local areas  as a function of their angular distribution (Fig.  \ref{fig:DM}B) and observe a larger fraction of successful detection on the transverse direction. This suggests that wavefront curvature and angle of incidence with respect to local portions of the cable \cite{fichtner2022} may have a role in building the cable response and highlights the importance of increasing the catalog size to allow for more conclusive claims.

\begin{figure}
    \centering
    \includegraphics[width=0.9\linewidth]{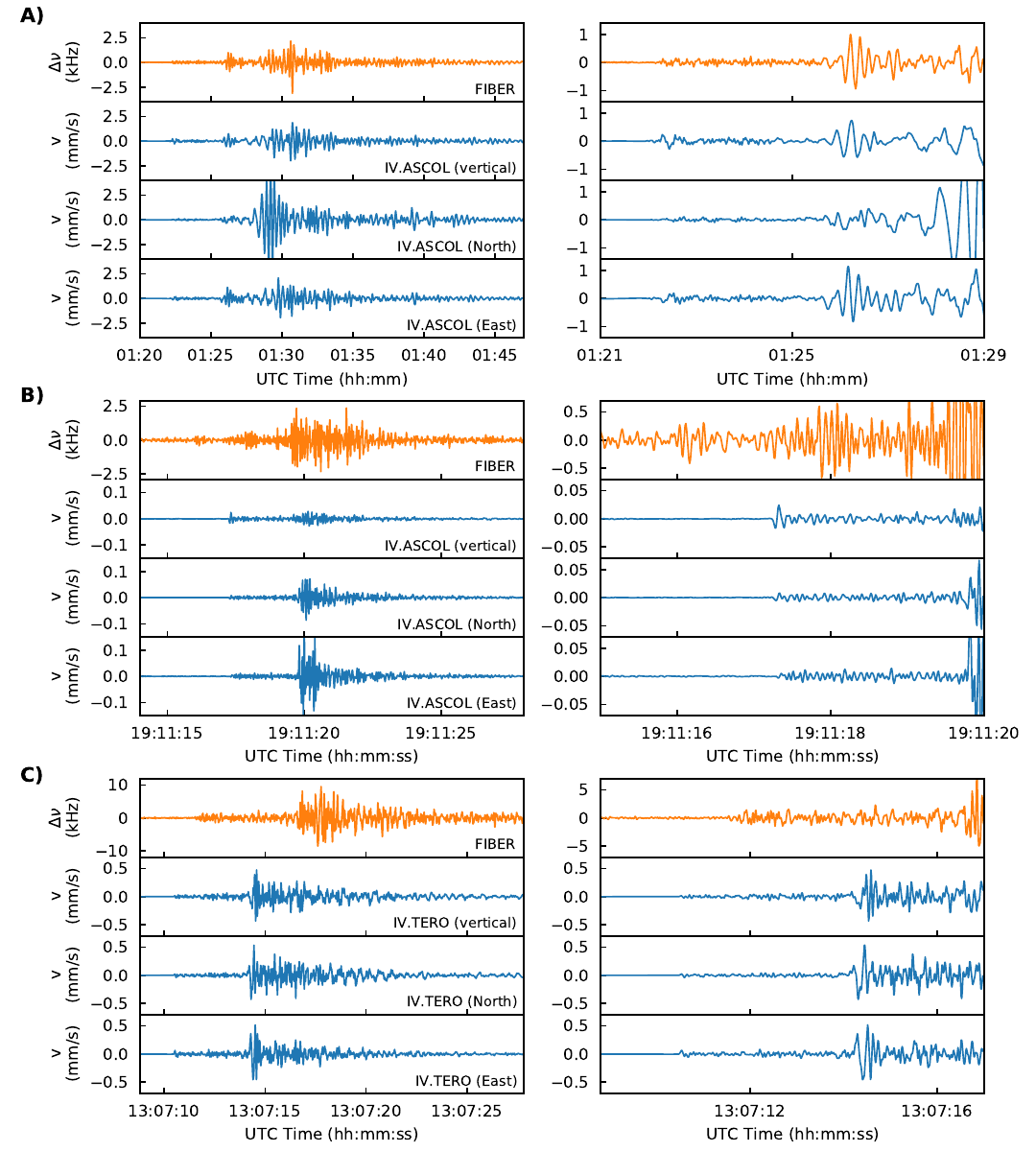}
    \caption{\textbf{Time traces of relevant case studies.} A) Recordings of the Turkey $M_{wp}$= 7.9 event occurred on Feb. 6th, 2023, with the fiber (panel 1, orange) and IV.ASCOL  (panels 2 to 4, blue). On the right, a zoom on the P and S-waves arrivals. 
    B) Recordings of the $M_{l}$=2 event occurred in Civitella del Tronto on Feb. 19th, 2022 with the fiber (orange) and IV.ASCOL (blue). C) Recordings of the $M_{l}$=3.4 event occurred in Accumoli on Feb. 27th, 2022 with the fiber (orange) and IV.TERO (blue).  
    Data from both the fiber and the seismometer have been band-pass filtered  with a 6th order, bidirectional Butterworth filter to maximize signal/noise ratio on the fiber recording. The pass-band was 
    \SI{0.005}{Hz} to \SI{0.5}{Hz} for the Turkey event, and  \SI{1.3}{Hz} to \SI{20}{Hz} for the Civitella and Accumoli events. Seismometers data were deconvolved from the sensor's response.}
    \label{fig:turchia_civitella_accumoli}
\end{figure}

For a direct comparison of fiber and  conventional seismometer measurements, an appropriate conversion metrics between quantities had to be considered, as recordings substantially differ from two points of view: first, fiber sensors measure an optical phase, while seismometric sensors measure ground velocities or accelerations; second, fiber sensor measurements are related to the integral of a distributed deformation, while seismometric sensors measure local quantities. Operationally, we record the time derivative of the  phase $\varphi(t)$ accumulated in a round-trip, i.e. the instantaneous frequency deviations of the returning optical signal with respect to the injected one $\Delta \nu (t) =\dot{\varphi}(t)/2\pi$. As other  authors \cite{bogris2022}, we found similarities in the time evolution and spectral composition of this quantity with the ground velocity $v(t)$ recorded by nearby seismometers, and understand them to be consistent with  expectations from a functional point of view (see Methods).
As an example, Figure \ref{fig:turchia_civitella_accumoli} compares $\Delta \nu (t)$ (orange, uppermost panel) with $v(t)$ measured by seismometers along the vertical, North-South and East-West axes (blue, panels 2 to 4) for the three events marked in Fig. \ref{fig:DM}. For each event, zoom on the P and S-wave arrivals is shown in the right panel.   

Figure \ref{fig:turchia_civitella_accumoli}A shows traces of the event occurred in Turkey on Feb. 6th, 2023  (moment magnitude 
 \cite{tsuboi1999} $M_{wp}$= 7.9, distance to the fiber \SI{2000}{km}, labelled 1 in Fig. \ref{fig:DM}).
The high signal/noise ratio on the fiber recording allows a sharp picking of the wave arrival time, with similar quality as obtained from seismometer data, and a clear identification of the S-wave arrival. We note that the scaling between the amplitude of the perturbation on the fiber and the vertical and North-South velocity components of IV.ASCOL is  about \SI{e6}{Hz/(m/s)}, which is also  in qualitative agreement with the value expected from a simplified configuration in which a straight fiber is solicited at one end \cite{kennett2022,fichtner2022,wang2018,bogris2022} (see Methods).
This scaling is confirmed for another far-field event, where correlation is also found between the time evolution of fiber and IV.TERO recordings (see Extended Data Fig. 5).

Figure \ref{fig:turchia_civitella_accumoli}B shows traces of an event occurred in Civitella del Tronto on Feb. 19th, 2022 ($M_l$=2, epicenter depth \SI{18}{km} and hypocenter distance  \SI{5}{km} from the closest point of the fiber, labelled 2 in Fig. \ref{fig:DM}).   Because of the proximity, the  seismic phases arrive earlier at the fiber location than at IV.ASCOL. However, the short distance to the cable may entangle the identification of phase arrival times, preventing from precise picking.
The scaling coefficient between recordings on the fiber and seismometer is about \SI{2e7}{Hz/(m/s)}, a factor $\sim$20 higher than for the Turkey event (this factor is consistent with other observations on a regional scale). This supports the  indication of Fig. \ref{fig:DM}B,  that the wavefront curvature and angle of incidence with respect to local portions of the cable \cite{fichtner2022} may have a non-trivial impact in determining the overall cable response and amplification. This condition is opposite to the previous case (far-field event), in which the seismic components reaching the cable had wavelength comparable to the cable length, thus washing out local effects and giving rise to a response which is more similar to that of a point-like sensor. More quantitative considerations would require integration of local strain, derived from a-priori knowledge of the wave propagation dynamics \cite{noe2023} which was not available for this event.

Fig. \ref{fig:turchia_civitella_accumoli}C shows the recordings of an event occurred in Accumoli ($M_l$=3.4,  epicentre at \SI{32}{km} from the fiber and \SI{27}{km} from IV.TERO, labelled 3 in Fig. \ref{fig:DM}).  The scaling factor between fiber and seismometer measurements is similar to the previous case. The arrival time of the P- and S- waves is evident, and the lag is consistent with the wave propagation between IV.TERO and the fiber to within the model uncertainty.  

These results confirm that  interferometric fiber recordings can be  used for earthquake localization based on the picking of phase arrival times on a set of cables \cite{marra2022}, provided that standard localization algorithms are adapted to take into account the distributed nature of the sensor. 

Spectral analysis discloses additional information about earthquakes dynamics. 
In standard seismometers, Fourier amplitude spectra of accelerations ($\dot{v}$) are predicted to  follow a standard Brune’s model \cite{brune1970}  and increase proportionally to $f^2$, with $f$ the Fourier frequency, up to a corner frequency $f_\text{c}$, above which they flatten. 
 The model also predicts a relation between $f_\text{c}$, the source physical parameters and the event magnitude (see Methods).  

\begin{figure}
\centering
    \includegraphics[width=0.8\textwidth]{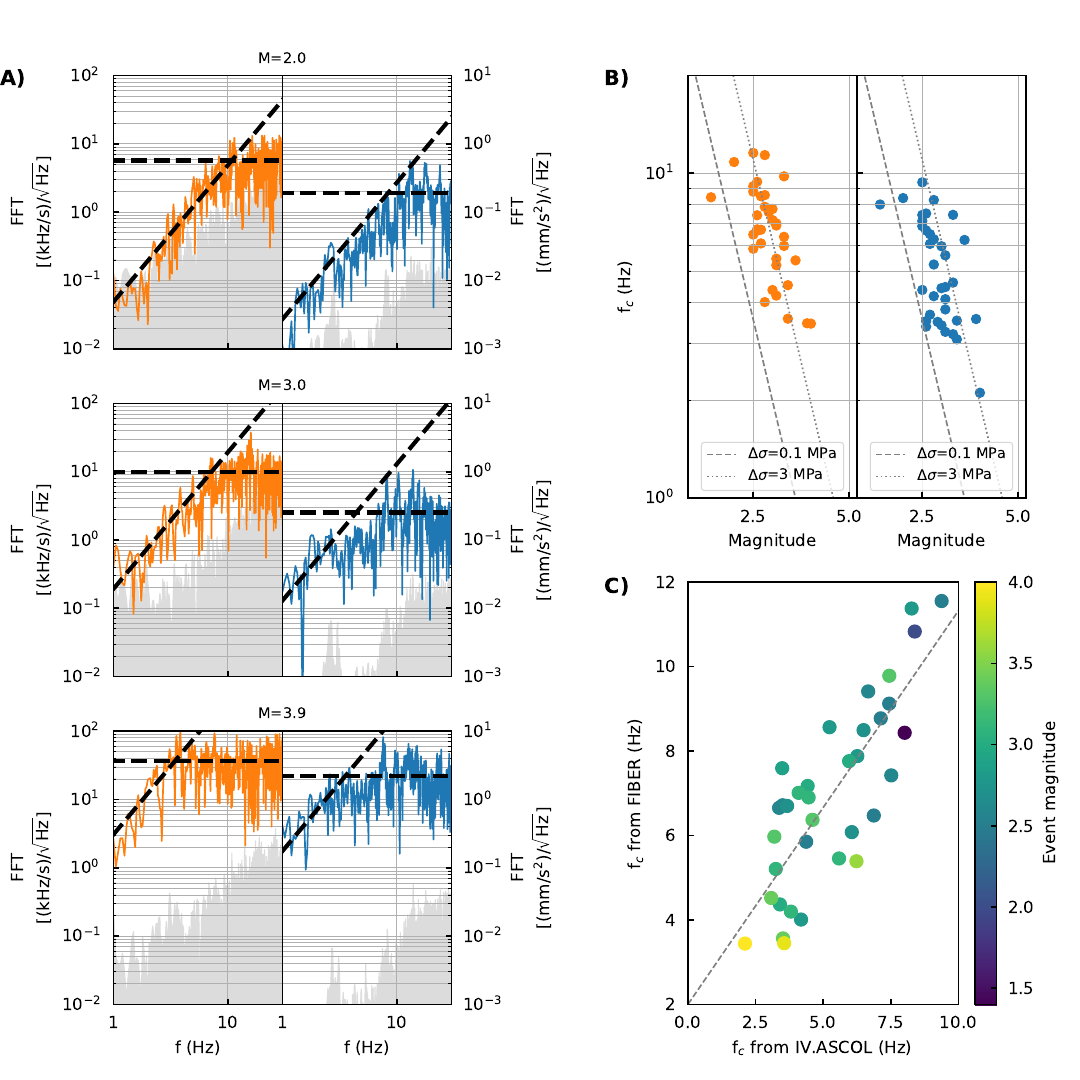}
    \caption{
    \textbf{Spectral analysis and magnitude determination.} A) Spectral composition of the accelerations recorded by IV.ASCOL East-West component (blue, right panel) and  fiber differentiated frequency data (orange, left) for three events with epicenters in Civitella (1st and 2nd row) and Accumoli (3rd row) and $M_l$=2.0, 3.0 and 3.9. The spectral response of the fiber well reproduces that of the seismometer.  Black dashed lines indicate fits on the low and high end of the detection band, their crossing corresponding to $f_\text{c}$. The shaded area indicates the background noise recorded by the sensor 1 minute before the event . B)  $f_\text{c}$ as retrieved from IV.ASCOL (right, blue) and fiber (left, orange) are plotted as a function of magnitude, evidencing a linear relation between the two quantities. Dashed and dotted lines indicate lower and upper limits obtained for typical source parameters; in particular, we let the stress drop $\Delta \sigma$ to vary between \SI{0.1}{MPa} and \SI{3}{MPa} (see Methods). C) Dispersion plot of the corner frequency as extrapolated from the fiber and IV.ASCOL, with color scale representing magnitude. Correlation between the response of the two sensors is found ($r=0.81$), suggesting that spectral analysis of  fiber recordings enables to infer event magnitudes. The dashed line is a linear fit to the data. }
\label{fig:spectra_main}
\end{figure}

We  analysed the spectral composition of all events detected by both the fiber and IV.ASCOL in a range of \SI{100}{km} from our observatory, and extrapolated the respective corner frequencies. Farther events are excluded, as their spectrum is expected to be significantly affected by propagation. As an example, spectra of three events with increasing magnitude are shown in Fig. \ref{fig:spectra_main}A (spectra of all events are shown in Extended Data Fig. 6).

The spectral content of fiber recordings reproduces that obtained by IV.ASCOL, highlighting the broadband response of the cable as a sensor and corroborating our decision to compare recordings of $\Delta \nu$ with $v$, or the corresponding time derivatives. Also, we note  that in all spectra the scaling between the cable and seismometer is consistent with that observed for the Civitella and Accumoli events (Fig. \ref{fig:turchia_civitella_accumoli}B and C), and generally higher than observed for the Turkey (Fig. \ref{fig:turchia_civitella_accumoli}A) event.
Figure \ref{fig:spectra_main}B shows the corner frequency values extrapolated from the fiber (orange, left) and IV.ASCOL (blue, right) as a function of magnitude, highlighting in both cases a decreasing trend consistent with the expected one. The Pearson's correlation coefficient between  magnitude and $f_\text{c}$ is $r=-0.60$  and  $r=-0.59$ 
respectively, indicating that the reason for statistical dispersion is not related to the data quality. Instead, we ascribe it  to differences in the physical source parameters between various events, as indicated by the predicted limits  when typical numbers  are plugged into the model (dotted and dashed lines, see Methods).  A Pearson's test between the corner frequencies retrieved by the two sensors (Fig. \ref{fig:spectra_main}C) returns $r=0.81$,  confirming a good correlation between the two. However, values extrapolated by the fiber appear to be systematically higher than for IV.ASCOL. This is confirmed by a  linear fit to the data in Fig. \ref{fig:spectra_main}C, which returns a slope of $(0.9 \pm 0.1)$\SI{}{Hz/Hz} and intercept of $(2.0 \pm 0.7)$ \SI{}{Hz}. Further inspection  from a larger event catalog  will help finding an explanation (see also Methods).

This analysis reveals a crucial aspect of fiber-based earthquake detection, i.e. that spectral analysis of fiber deformations can indeed support magnitude estimation for  small and mid-size seismic events, even in the impossibility to calibrate the absolute cable response to ground motion due, e.g., to unknown fiber-to-ground coupling parameters or integration effects.

 \section*{Discussion}
With 1.5 years of uninterrupted data collection on a land-based fiber shared with data traffic, we were able to characterize laser interferometry as a tool for earthquake detection, also taking advantage from the presence of co-located seismometers with well-known response. 
Within the same period, no degradation in data traffic quality metrics was reported, which is a prerequisite for shared use of the existing fiber infrastructure.

 We showed that events with $M_l$=2 or larger can be reliably detected at local distances. For the Italian earthquake catalogue  \cite{iside}, $M_l$=2 represents the average magnitude for
completeness and in the surveillance system the alert-threshold to the Civil Protection Department  is $M_\text{l}$= 2.5  \cite{margheriti2021}: smaller events are not felt unless they have very superficial hypocenters or happen in quietest moments of the day.
  Detection of such events with a fiber hosted in a road infrastructure, extending to city centers  and including aerial segments reveals  concrete potential of laser interferometry for supporting traditional monitoring in highly-populated areas. It also highlights the capability to bring data from weak earthquakes that might not be traced unless seismometers are present in a range of few tens of kilometers from the epicentre. Such a dense coverage is in general only available in high-risk areas and is uncommon on the broadest extent of the planet. Considering the current increase in fiber cable deployment to meet the growing capacity needs for next-era communications, we believe the interesting target for network-integrated  sensing to be weak yet close events, rather than teleseismic signals for which the existing network is already adequate. 
In the perspective of integrating laser interferometry into telecommunication grids, the scaling law we found for the event detection probability can be exploited to draw territorial  maps of  covered regions by following the hosting infrastructures path and validate small-event detections in combination with spectral analysis, even lacking information on detailed cable routes and mechanical coupling to the ground. 

Sharply measuring the arrival time of the P and S seismic phases is a fundamental step for  localization.
We provided examples of this capability for events diverse in magnitude and distance, showing that event picking is possible, especially with the S-wave often featuring a steep and distinguishable onset.

Availability of large catalogs from land-based observatories, validated by comparison with seismometer data,    is crucial in developing advanced signal analysis and training dedicated machine learning routines \cite{zhu2008} and allows studying the fiber sensitivity to ultra-low-frequency signals. Clear signature of daily temperature variations was observed on fiber recordings, which could be compared in the future to other distributed temperature sensing methods or extended to other slow-varying phenomena as suggested in \cite{marra2022}.

From another perspective, the analysis of background noise confirms that this technology is suitable for monitoring a large variety of phenomena besides earthquakes, such as vehicle traffic and infrastructure mechanical resonances.

For our experiment, we made use of a research-grade, sub-Hz linewidth laser, although results indicate that performances can be relaxed in favour of integration and lower cost, competitive to those of other sensors.
Notably, phase detection could be integrated into modern transceivers \cite{ip2021} similarly to what demonstrated for polarization-based sensing \cite{zhan2021}.  Although phase analysis requires more stable lasers than used today in telecommunications, laser  integration technologies are progressing \cite{kelleher2023}; moreover, thanks to the linear relation with the fiber strain, it features lower computational requirements and higher sensitivity than polarization analysis \cite{ip2021}. Overall, this may represent in the future the most effective route to the integration in modern smart multi-service grids in a scalable and sustainable approach.

\bibliography{Bibliography}
\bibliographystyle{sn-nature}

\section*{Methods}
\subsection*{Experimental layout}
The fiber optical length is sensed using an inteferometric scheme, where an ultrastable laser beam is split into two branches, one of which travels the link in a round-trip before being recombined to the other. The phase difference between the two arms is attributed to combined variations of the  refractive index and length of the deployed link, induced by changing strain in the surrounding environment. 
A fixed frequency shifter  is introduced to enable detection in the radio-frequency domain instead of the low-frequency band, where detection noise would be degraded by the sampling electronics. The beatnote between the round-trip and the reference beam  is  demodulated, sampled and processed to extract the optical carrier frequency variations at a typical rate of \SI{1}{kHz}, then stored in a database. Collected data are filtered and decimated to a final rate of \SI{100}{Hz} for analysis. 
The instrumentation is housed at the telecom network nodes and installed in compliance to rack standards. Part of the measurements were collected with a modified optical scheme, based on a two-way transmission. 
In this configuration, a pair of lasers, each housed in a network node, is launched towards the remote end and interfered with the local laser. The frequency shift experienced by the laser while travelling through the fibers is extracted by combining the interference signals recorded at the two cable ends \cite{donadello2023}. 
In both schemes, we make the assumption that perturbations are  symmetrical on the two directions, as the fibers are laid parallel into the same cable. 

Equipment at the two nodes is hardware-synchronized to  \SI{5}{\micro s}, and referenced to Universal Coordinated Time (UTC) to about \SI{10}{ms} using 
Network Time Protocol provided by Ethernet connection.

The link implements Dense Wavelength Division Multiplexing (DWDM). All optical carriers, included the one used for this work, are combined through an Optical Add/Drop Multiplexer, equalized and launched into the fiber. An optical preamplifier is installed at the far cable end, before de-multiplexing, to compensate for a \SI{10}{dB} optical loss. Launching from the opposite direction follows the same scheme. 
The ultrastable laser has been installed into the DWDM hardware as an alien lambda with the possibility to equalize and raise alarms in case of faults to the system.

\subsection*{Seismic stations}
Data acquired from the optical fiber were compared with seismological data acquired by IV.TERO and IV.ASCOL seismic stations. 
IV.TERO is the seismic station of the Italian National Seismic Network \cite{ivnetwork,margheriti2021}) closest to the fiber
end in Teramo, Italy. The station is located at  670 metres above mean sea level, along the slope of a headland in the piedmont of the Central Apennines.  IV.TERO is equipped with velocimetric (Nanometrics TRILLIUM-40s) and accelerometric (Kinemetrics EPISENSOR-FBA) sensors, connected to a 6 channel GAIA-2 seismic digitizer. All instruments are located in an inspection pit approximately 2 m deep from the ground level. The geophysical and geological surveys performed for the site characterization of the station indicate a stiff site characterized by a value of VS30 of 981 m/s.

The station IV.ASCOL was installed in the town of Ascoli, on the ground floor of the network point-of-presence that houses the laser interrogator. 
The station consists of a velocimetric (Lennartz Le3D-5s) and an accelerometric sensor (Kinemetrics EPISENSOR-FBA) connected to a 6 channel REFTEK-RT130 seismic digitizer. The hardware was lent by the COREMO program of INGV especially for this experiment. 
Although located in an unfavorable area,  characterized by high anthropogenic noise and possible site amplification phenomena (the site is on fluvial deposits), it was decided to install the station in close proximity to the fiber cable, for  easier comparison of the recordings.
Both IV.TERO and IV.ASCOL stations send, by WIFI and UMTS transmission respectively, acquired data in near-real-time to INGV's data collection server, where they can be gathered and distributed via Web Services based on the FDSN specification \cite{webpageingv}.

\subsection*{Selection of events for the sensitivity analysis and validation criteria}

From continuous acquisitions, we automatically extracted windows where to look for signature of seismic events according to the expected arrival times of the seismic wave  to  fiber, IV.ASCOL and IV.TERO, derived from a-priori knowledge of the event source position and propagation parameters. 
For each earthquake, the used source parameters (origin time, hypocenter coordinates and magnitude) were extracted from the seismic catalog produced by INGV seismic surveillance center \cite{surveillance_ingv}. The source-fiber distance was computed with respect to the middle of the cable (\ang{42.73693}N, \ang{13.674261}E). 

Waveforms of earthquakes recorded along the fiber and at seismic stations were selected using the theoretical P arrival times computed using the TauP method \cite{crotwell1999}. The travel times were computed in the 1D Earth reference velocity model ak135 \cite{kennett1995}, that provides a good fit to a wide variety of seismic phases and improves on the previous IASP91 \cite{kennett1991} Earth model in terms of  data and methodologies.

We considered in our analysis events with: $M \ge 4$ closer than \SI{2000}{km}, or $M_l \ge 2.5$ closer than \SI{300}{km}, or $M_l \ge 0.4$ closer than \SI{30}{km}, occurred between June 19th, 2021 and Sept. 26th, 2022. To increase statistics in the high-distance/high-magnitude end of the plot, we added events from the recent sequence occurred in Turkey, including those with $M_{wp}\ge$ 5 occurred between Feb. 6th and March 23th 2023 closer than \SI{300}{km} to the epicenter of the first event. Also, we included a seismic sequence occurred off-shore Ancona, at a distance of about \SI{130}{km} from the middle of the fiber (coordinates of main earthquake of the sequence, Mw 5.5: \ang{43.9833}N, \ang{13.3237}E) in November 2022, that featured co-localised events in a broad range of magnitudes. To include this sequence,  we considered events with $M_l \ge 2$ closer than \SI{180}{km} from the fiber, occurred between Nov. 9th, 2022 and Nov. 18th, 2022.  

For each considered timeframe, we looked for signature of the wave detection on the fiber according to typical criteria used in seismic analysis and assigned it a score (0 = event not detected; 100 = event clearly visible; intermediate values reflect the confidence on detection and are plotted in light-green color in Fig. 2 of the main text). 
To make a decision, we considered both the time trace (i.e. clear onset of a phase from background noise; temporal evolution and duration in consideration of epicenter distance; matching of the recording with the expected arrival time of the seismic wave...) and the spectral composition of the signal.

 In Fig. 2 of the main text, events marked  green are those who featured a score $\ge$ 60; unsure events are those who featured a score between 5 and 60.

\subsection*{Interpretation of fiber data and comparison to seismometer recordings}

The phase accumulated by the optical carrier as it travels a fiber with length $L$ can be related to the integral of local ground strain $\epsilon$ along the full cable path, weighted by  a geometrical scaling coefficient $\gamma_\text{g}$ and a coupling coefficient $\gamma_\text{c}$. The former depends on the  relative orientation of the cable and the local strain direction \cite{fichtner2022, kennett2022}  and on the ratio between the seismic wavelength and the fiber length projected along the wave direction: for instance, because of the integration effect, we would expect null response from a perfectly straight cable hit by a seismic wave incident from the longitudinal direction when the cable ends happen to be in nodal points; instead, response would be higher when the cable ends happen to be perfectly out of phase  \cite{kennett2022}. From a practical point of view, this sort of considerations, well-modeled in standard Distributed Acoustic Sensing,  cannot be extended to coherent interferometry in a straightforward way, as real cables always feature changing orientation and have dimensions comparable or larger than seismic wavelengths \cite{fichtner2022}.
$\gamma_\text{c}$ quantifies the anchoring of the fiber to the surrounding ground and depends on construction factors such as the kind of conduit where the fiber is placed, the cable armoring \cite{ajo2019,flores2023} or the presence of gel rather than simple air-filling \cite{follett2014}. This establishes a general relation between the instantaneous frequency deviations of the optical carrier with respect to the nominal frequency of the injected signal $\Delta \nu = \dot{\varphi}/2\pi$ , and the  time derivative of integral strain:
\begin{equation}
\label{eq:fib_strain}
    \Delta \nu = \frac{1}{2\pi} \dot{\varphi}=  2\frac{nL}{\lambda}\int_L \gamma_\text{g}(s)\gamma_\text{c}(s) \dot{\epsilon} \, ds
\end{equation}
where we parametrized $\gamma_\text{g}$  and $\gamma_\text{c}$ accounting for the fact that they are local quantities, $\lambda$ is the optical wavelength (\SI{1.5}{\micro m}), the factor 2 indicates that the fiber is travelled in a round-trip  and we assumed a constant refractive index $n$ throughout the cable. 
The relation between strain, or its derivative, and ground motion parameters is, in general, less trivial. Observations with standard Distributed Acoustic Sensing confirm that a linear relation exists between $\dot{\epsilon}$ and ground velocity components $v$ along the strain direction recorded at the edges of the gauge length $g$ \cite{kennett2022, wang2018,zhu2021}, with $g$ of the order of few  meters: 
\begin{equation}
\label{eq:strain_gauge}
    \dot{\epsilon}=[v(g/2)-v(-g/2)]/g
\end{equation} 
However, this relation could be extended to the km-scale only in the  case of a perfectly straight fiber under  homogeneous strain conditions.
On deployed fiber layouts with changing orientation and non-homogeneous deformations, knowledge of velocity components at the cable end are not sufficient to quantitatively predict the cable response and propagation models  for the seismic wave need to be computed to enable integration of local deformation through  the cable path \cite{noe2023}.

Still, we expect a linear functional relation between the recorded frequency deviations and the ground velocity to be preserved, which is consistent with the reported measurements for the analysed cases. 

\subsection*{Calculation of spectra and corner frequency extrapolation}
For all detected events in a range of \SI{100}{km} from the the fiber we computed the fast Fourier transform (FFT) of the recording over a time interval of \SI{20}{s}, with Hann window and no averaging to enable sufficient resolution at lowest frequencies. 
The start time of signal windows was automatically set to be \SI{1}{s} before the calculated arrival time of the P-wave in the center of the cable, except for two cases where it had to be fixed manually due to incorrect arrival time prediction.
Similar procedure was followed to calculate the background noise (shaded area): integration intervals had the same duration, and the start time was either automatically set to \SI{60}{s} before that of the signal or manually adapted to exclude non-stationary noise arising by chance during the considered period. In this latter case we shifted the window by up to few tens of seconds,  maintaining the same duration. 
The modulus of the FFT was normalized by the square-rooth of the channel bandwidth. Results are shown in Extended Data Fig. 6 both for the fiber  (orange) and IV.ASCOL East-West component (blue).

To extrapolate the corner frequency we fitted the above spectra with polynomials of the kind $\text{FFT}(f)=b_2f^2$ and $\text{FFT}(f)=b_0f^0$ in Fourier frequency intervals \SI{1.2}{Hz}-\SI{3}{Hz} and \SI{10}{Hz}-\SI{30}{Hz} respectively. The interpolation limits have been chosen to exclude spectral regions where the noise on fiber recordings covers the signal, and especially the discrete noise peaks at \SI{0.75}{Hz} and \SI{0.95}{Hz}. The corner frequency is calculated as the crossing point between the two fitted lines, i.e.  $f_\text{c} = \sqrt{b_0/b_2}$. 

The corner frequency is related to the source physical parameters and seismic moment $M_\text{o}$ according to \cite{morasca2022}:
\begin{equation}
    \label{eq:cornerfreqMo}
    f_\text{c} =kC_\text{d} \beta \left(\frac{16}{7} \frac{\Delta \sigma}{M_\text{o}}\right)^{1/3}
\end{equation}
with $\Delta \sigma$ the stress-drop, $C_\text{d}$ the directionality coefficient ranging between 0.1 and 10, $\beta$ the rupture velocity and coefficient $k=0.37$.
Using the relation $M_\text{o}=10^{1.5M+9.1}$ \cite{hanks1979}, with $M$ the event magnitude,  we can derive the relation between $f_\text{c}$ and $M$:
\begin{equation}
    \label{eq:cornerfreq}
    f_\text{c} =kC_\text{d} \beta \left(\frac{16}{7} \frac{\Delta \sigma}{10^{1.5M+9.1}} \right)^{1/3}
\end{equation}
 
Clearly, all these parameters may change from an event to the other, although the greatest variability is found for $\Delta \sigma$, that may vary  on almost two orders of magnitude in the range \SIrange{0.1}{3}{MPa}.

The dashed and dotted lines shown in Figure 4B of main text indicate the corner frequency extrapolated from Eq. \ref{eq:cornerfreq} using $\beta$=\SI{3}{km/s}, $C_\text{d}$=1 and $\Delta \sigma$= \SI{0.1}{MPa} or \SI{3}{MPa} respectively. A decreasing trend of the corner frequency is observed as the event  magnitude increases, with the dispersion of results consistent with inhomogeneities in source parameters. 

In our derivation of corner frequencies, we did not take into account propagation effects on the source spectrum, and did not apply any threshold on the event detection quality, based e.g. on signal/noise considerations. These effects may also play a role in explaining the dispersion of experimental results.
As highlighted in the Main text, some bias is clearly distinguishable between values derived from the fiber or IV.ASCOL, whose reason deserves further investigation. At this stage, we cannot exclude some bias in the fit of the low-frequency components of spectra with a polynomial of the kind $b_2f^2$ component, due to the high noise at \SI{0.75}{Hz} and \SI{0.95}{Hz} affecting the cable. 

The analysis was repeated with the vertical and North-South components of IV.ASCOL recordings with no substantial differences.

\subsection*{Data and materials availability}
Raw traces that were used to produce the images appearing in the paper are publicly available \cite{zenodolink}; integral datasets and the full updated catalogues are available upon request to the authors.




\section*{Acknowledgments}

We thank the COREMO program from INGV for the loan of a seismic station and the technical staff of INGV for the maintenance of the seismic network, including IV.TERO.

The project was funded by Open Fiber and Metallurgica Bresciana within the MEGLIO project "Monitoring of Earthquake signals Gathered with Laser Interferometry on Optic fibers" and by the European Metrology Program for Innovation and Research (EMPIR) project 20FUN08 NEXTLASERS. The EMPIR initiative is co-funded by the European Union’s Horizon 2020 research and innovation programme and the EMPIR Participating States.

\subsection*{Author contributions}
S. D. and C. C. designed, realized and operated the optics and acquisition  apparatus.
A. G., A. H., L. M., M. V. provided seismological expertise and processed seismological events.
M. H., D. B. and F. C. provided telecom network expertise and managed the integration of the interferometric signal  to the telecommunication infrastructure from the architectural, hardware and logistical point of view.  
R. C., E. B. and A. M. developed the fiber sensor electronics and contributed to the sensor installation.
F. L., A. M. and D. C. provided optics expertise and background knowledge on the interferometric sensor design.
S. D., A. G., A. H., M. V., C. C., L. M. processed the data, performed the technical analysis and wrote the paper, with contributions from all authors.
D. C., F. C. and A. H. coordinated the project.
\subsection*{Competing Interests}
The authors declare no competing interests. 
\setcounter{figure}{0}
\renewcommand{\figurename}{\textbf{Extended Data Fig.}}
\renewcommand{\thefigure}{\textbf{\arabic{figure}}}

\begin{figure}
\centering
\includegraphics[width=0.5\linewidth]{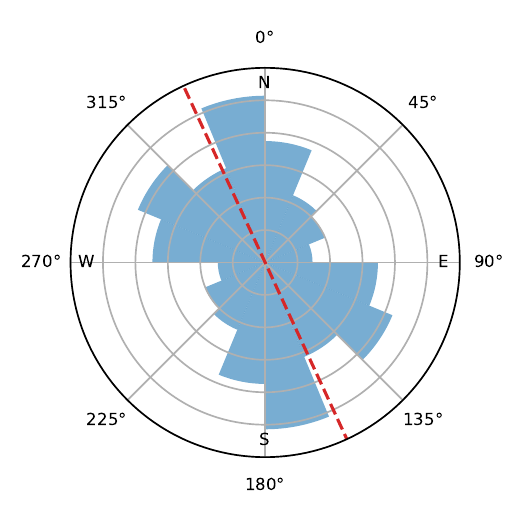}
\caption{\textbf{Average azimuth of the cable.} We approximate the full cable path to a series of straight segments and for each we calculated the orientation. Polar histogram bars indicate most recurring orientation angles. The red line is the result of a circular mean of all segments and indicates that the cable has an average azimuth of \ang{155}. \label{fig:fiber-distrib}
 }
\end{figure}
\begin{figure}
\centering
\includegraphics[width=0.9\textwidth]{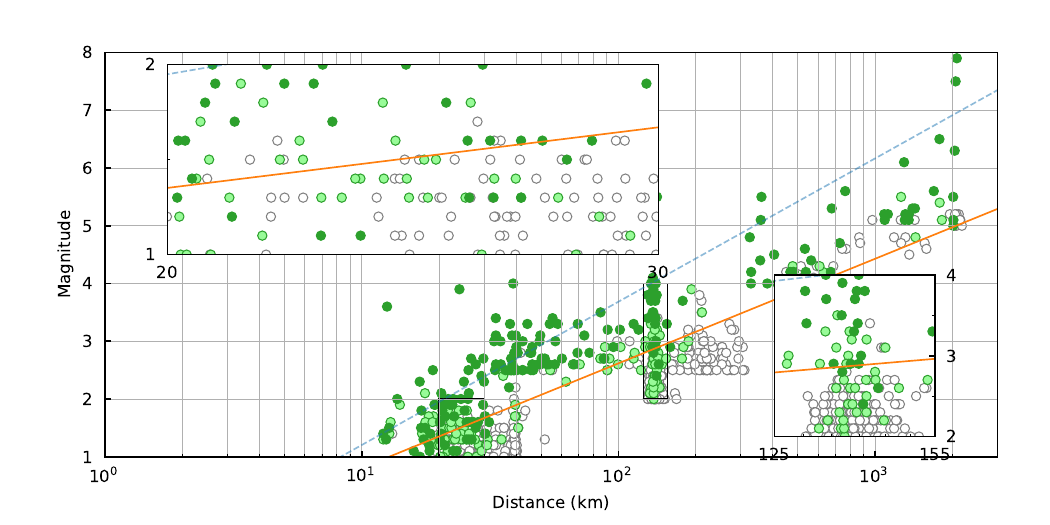}
    \caption{\textbf{Sensitivity of IV.ASCOL.} Map of events detected (green),  unsure (light-green) and non-detected (empty) by   IV.ASCOL, as a function of their magnitude and distance to the sensor; validation criteria were the same as for the fiber. The dashed blue line reproduces the sensitivity threshold of the fiber, shown in the main text; the solid orange line is the one obtained for IV.ASCOL and corresponds to $A_1 = 0.552 \pm 0.001$ and $A_2=(3590 \pm 10)$ \SI{}{km}. The sensor suffers from the high amount of anthropogenic noise in the surroundings and has a higher average noise level as compared to other stations of the INGV network. In spite of this, IV.ASCOL is used in the INGV daily routine localization task, for instance during the Nov. 2022 sequence off-shore Ancona  (coordinates of main earthquake of the sequence, $M_{wp}$ 5.5: \ang{43.9833}N, \ang{13.3237}E).  The insets zoom on regions with highest density of events around the threshold.}
    \label{fig:DM_ASCOL}
\end{figure}

\begin{figure}
    \centering
\includegraphics[width=0.7\textwidth]{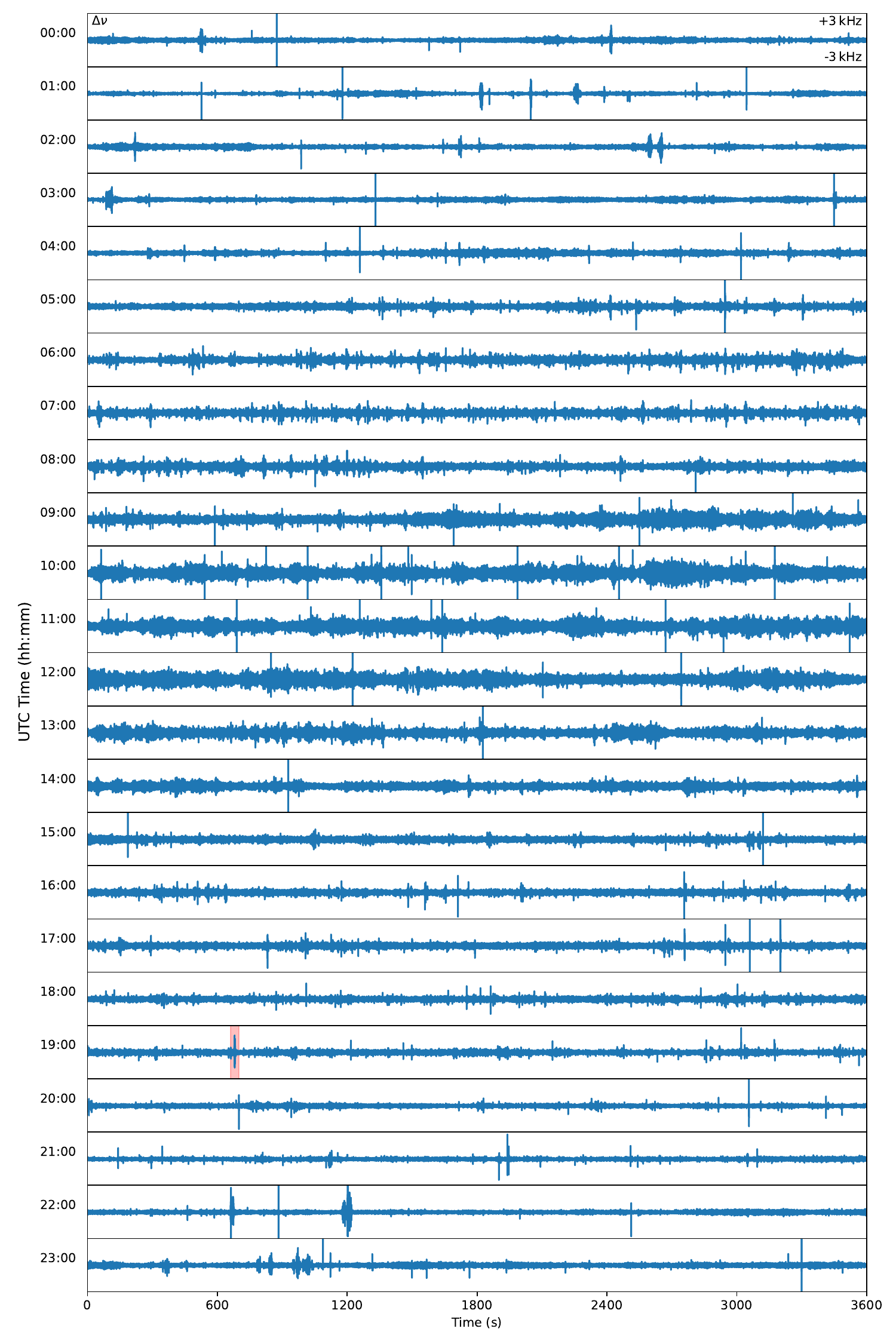}
    \caption{\textbf{Background noise of the fiber. }Time series of the unfiltered fiber recording over one day (one hour per line), showing increase of the noise at daily hours. We attribute this behaviour to higher vehicle traffic levels on the road where the fiber is hosted, and people activities around the cables, likely in city areas where the cable ends. Occasional, non-stationary solicitations to the cable are also visible. The red area corresponds to the Civitella event described in the main text.}
    \label{fig:background_one_day}
\end{figure}

\begin{figure}
    \centering
\includegraphics[width=0.7\textwidth]{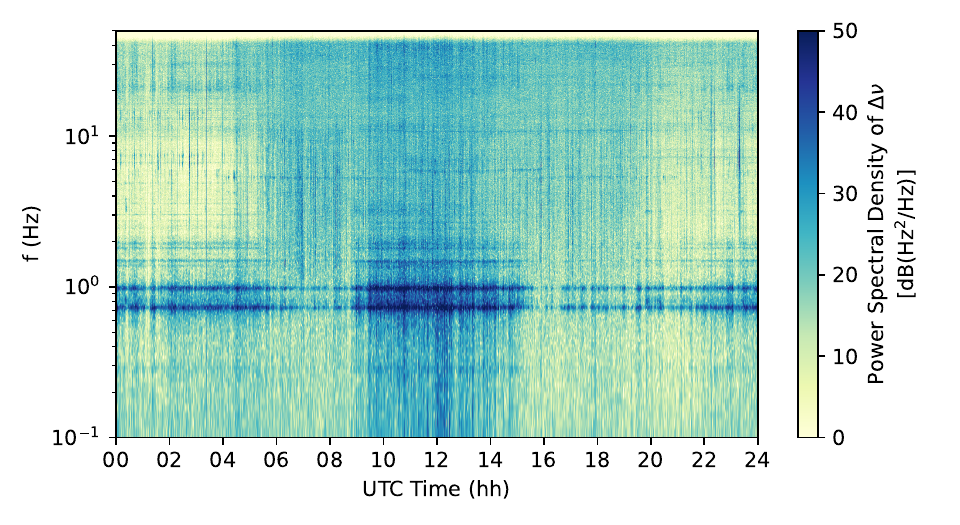}
    \caption{\textbf{Background noise of the fiber in the spectral domain. }Spectrogram of background noise on the fiber calculated over 24 h. More than \SI{10}{dB} variation is observed on the noise between day and night hours, as well as two narrowband disturbances at \SI{0.75}{Hz} and \SI{0.95}{Hz}, which we attribute to oscillation modes on a pair of suspended fiber spans in proximity of a \SI{95}{m} long bridge. The buildings close to the cable are not tall enough to explain  resonances at
these frequencies, while  systematic seismic measurements on all the bridges  along the route showed resonances all above \SI{1}{Hz}.}
    \label{fig:fibernoise}
\end{figure}

 \begin{figure}
     \centering
    \includegraphics[width=\textwidth]{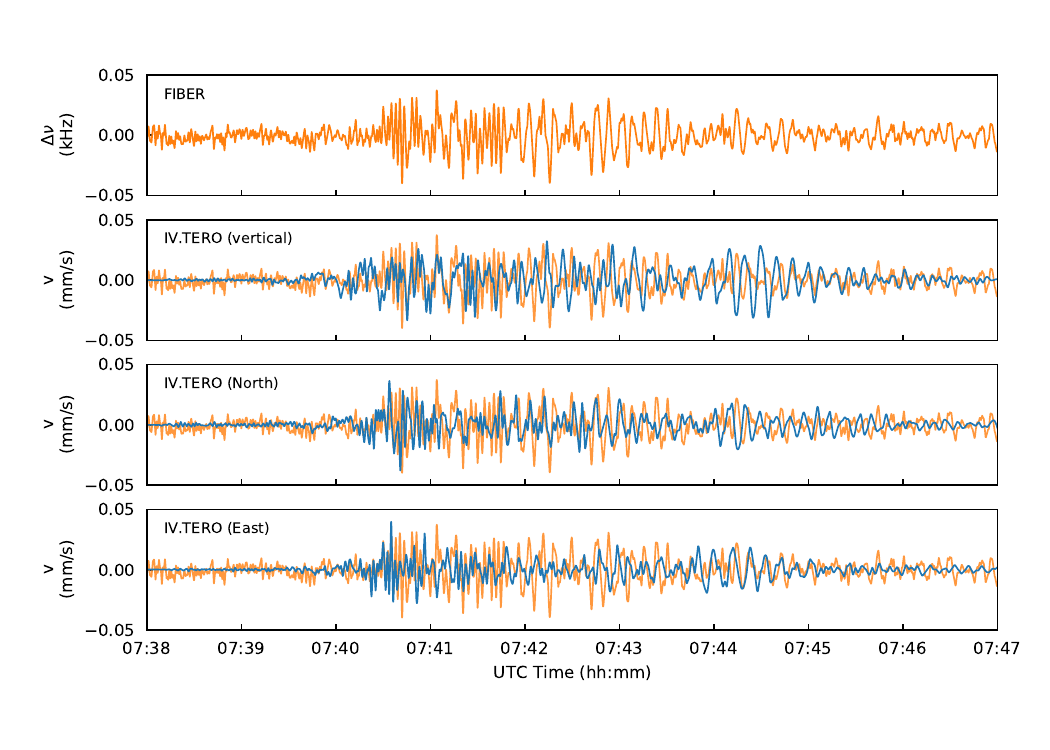}
     \caption{\textbf{Detection of an event in the Ionian Sea.}  Traces of the  M$_{wp}$=5.6 earthquake occurred on Sept. 8th, 2022  in the Ionian Sea, distance to the fiber \SI{765}{km}. The fiber recording is shown in orange, first panel; IV.TERO recordings along the three axes are shown in blue, panels 2 to 4 after deconvolving from the sensor response. 
    The fiber recording is divided by an arbitrary coefficient \SI{e6}{Hz/(m/s)} and overlapped to seismometer data. Qualitative agreement between the time evolution of the fiber and seismometer east-west
components is found at the trailing edge (07:40:30 UTC) and on the tail (07:44:30 UTC) of the event. Data have been filtered on a bandwidth \SI{0.005}{Hz} to \SI{0.5}{Hz} with a 6th order bidirectional Butterworth filter to maximize signal/noise ratio on the fiber recording. We attribute these similarities to the fact that, for far-field events, the cable behaves as a point-like sensor as the arriving seismic components have wavelength comparable to its length. }
     \label{fig:ionio}
 \end{figure}

\begin{figure}
    \centering
\includegraphics[width=0.8\textwidth]{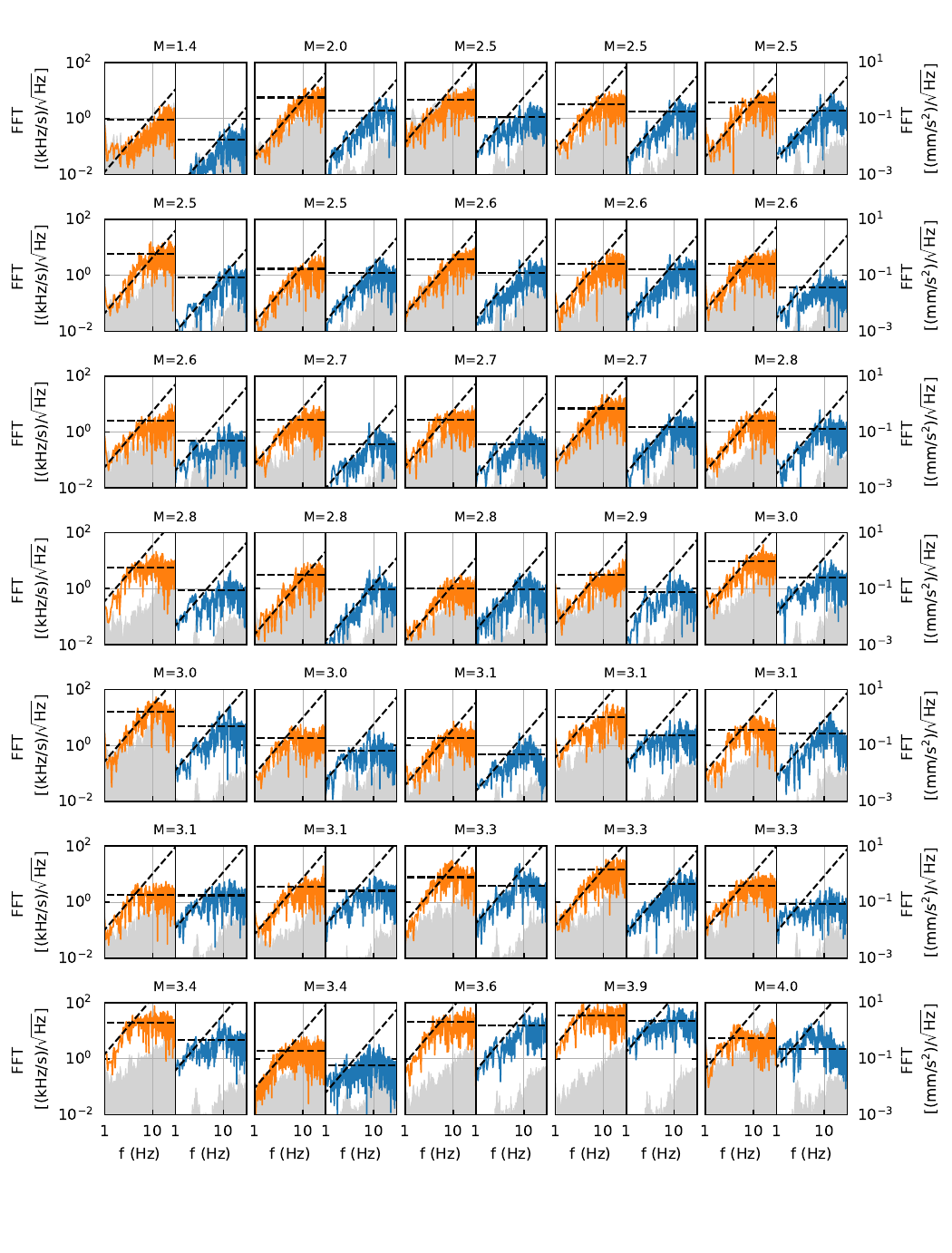}
    \caption{\textbf{Spectral analysis and corner frequency evaluation.} Fourier transform of east-west acceleration of IV.ASCOL (blue, right panel for each event) and differentiated fiber frequency data (orange, left). Gray area indicates background noise 1 minute before the event. Ticks and labels are indicated on outer axes, and are consistent between all traces. Black dashed lines are fits of the spectral distributions, their crossing corresponding to $f_\text{c}$. For the analysis,  we considered all detected events in a range of \SI{100}{km} from the fiber. IV.ASCOL data were deconvoluted from the seismometer response. }
    \label{fig:spectra}
\end{figure}

\end{document}